# The Role of Interdiffusion and Spatial Confinement in the Formation of Resonant Raman Spectra of Ge/Si(100) Heterostructures with Quantum-Dot Arrays

I. V. Kucherenko[a], V. S. Vinogradov[a], N. N. Mel'nik[a], L. V. Arapkina[b], V. A. Chapnin[b], K. V. Chizh[b], and V. A. Yuryev[b]

[a] *Lebedev Physical Institute, Russian Academy of Sciences, Leninskiœ pr. 53, Moscow, 119991 Russia*
*e-mail: kucheren@sci.lebedev.ru*

[b] *Prokhorov General Physics Institute, Russian Academy of Sciences, ul. Vavilova 38, Moscow, 119991 Russia*

**Abstract**—The phonon modes of self-assembled Ge/Si quantum dots grown by molecular-beam epitaxy in an apparatus integrated with a chamber of the scanning tunneling microscope into a single high-vacuum system are investigated using Raman spectroscopy. It is revealed that the Ge–Ge and Si–Ge vibrational modes are considerably enhanced upon excitation of excitons between the valence band $\Lambda_3$ and the conduction band $\Lambda_1$ (the $E_1$ and $E_1 + \Delta_1$ transitions). This makes it possible to observe the Raman spectrum of very small amounts of germanium, such as one layer of quantum dots with a germanium layer thickness of $\approx 10$ Å. The enhancement of these modes suggests a strong electron–phonon interaction of the vibrational modes with the $E_1$ and $E_1 + \Delta_1$ excitons in the quantum dot. It is demonstrated that the frequency of the Ge–Ge mode decreases by 10 cm$^{-1}$ with a decrease in the thickness of the Ge layer from 10 to 6 Å due to the spatial-confinement effect. The optimum thickness of the Ge layer for which the size dispersion of quantum dots is minimum is determined.

## 1. INTRODUCTION

In recent years, semiconductor nanostructures with quantum dots (QDs) have attracted the particular interest of researchers. The investigation of these nanostructures is of considerable importance for the understanding of the physics of low-dimensional structures. The compatibility of Ge/Si structures grown on silicon substrates with the well-developed silicon technology makes them attractive for fabricating optoelectronic and microelectronic devices. The design of devices with good parameters requires the knowledge of optical and electrical properties of these structures. A decrease in the size of quantum dots brings about a change in their electronic band structure and, as a consequence, a substantial increase in the efficiency of optical transitions [1]. In particular, Peng et al. [2] demonstrated that the intensity of photoluminescence from these quantum dots is considerably higher than the intensity of photoluminescence from quantum wells. At present, the influence of different parameters of quantum dots, such as the internal elastic stress, the size, and the composition, on their optical properties has been studied extensively. Raman spectroscopy is a powerful method for investigating structural properties of nanoobjects. The position and width of peaks in Raman spectra allow one to judge the stresses arising in layers, the interdiffusion of components, and the uniformity of the size distribution of quantum dots. Brya [3] and Renucci et al. [4] reported on the Ge–Ge, Ge–Si, and Si–Si vibrational modes in $Ge_xSi_{1-x}$ bulk alloys. The frequencies of these modes in Ge/Si nanostructures with quantum dots were studied in [5–7]. Self-assembled Ge quantum dots are formed as a result of lateral compressive stresses arising from the mismatch of the lattice parameters of the silicon substrate and germanium layers. The mismatch of the lattice parameters is equal to 3.8%. The shift of the peak of the Ge–Ge and Ge–Si modes enables one to judge the degree of stress in the layers.

In this work, the frequencies and widths of the lines associated with the Ge–Ge and Ge–Si vibrational modes in quantum dots were investigated as a function of the quantum-dot size by Raman spectroscopy. The study of the influence of Ge wetting layers and Si capped layer (spacer) on the Raman spectra was of special interest. In our samples, the thickness of Ge layers was varied from 6 to 18 Å.

## 2. SAMPLE PREPARATION AND EXPERIMENTAL TECHNIQUE

Structures containing Ge quantum dots were grown on a Riber EVA32 molecular-beam epitaxy apparatus integrated with a chamber of a GPI-300 scanning tunneling microscope (STM) into a single ultrahigh-vacuum system. In the course of experiments with the use of the above apparatus, the samples remain under ultrahigh-vacuum conditions (i.e., their surface is not subjected to contamination and oxidation) and can be transferred a required number of times to the chamber of the scanning tunneling microscope for investigation and back to the molecular-beam epitaxy chamber for further treatment and growth of new epitaxial layers. Since the samples during the experiment are not contaminated and oxidized, they can be studied with an atomic resolution at any stage of growth of the epitaxial heterostructure.

In this study, $p$-Si wafers with the (100) orientation, which were grown by the Czochralski technique and doped with bromine to a resistivity of 12 Ω cm (KDB-12), were used to grow structures.

Samples for Raman spectroscopic investigations were prepared as follows. After washing and chemical treatment, the initial wafers were subjected to preliminary heat treatment at a temperature of 590°C for 6 h in a chamber of preliminary annealing under high-vacuum conditions (residual pressure $\sim 5 \times 10^{-9}$ Torr). The natural oxide film was removed in the molecular-beam epitaxy chamber preliminarily evacuated to a residual pressure of $\sim 10^{-11}$ Torr. In order to remove the oxide film, the wafers were annealed at a temperature of 800°C under simultaneous irradiation of the operating surface of the wafer by a weak beam of silicon atoms. The deposition rate of silicon atoms during the deoxidization of the surface did not exceed 0.01 nm/s. After removal of the oxide film, an undoped silicon buffer layer ~100 nm thick was grown on the wafer surface (growth temperature, 550°C). Then, germanium quantum dots were grown on the buffer layer at a temperature of 350°C. When growing the layers with quantum dots, the effective thickness of the deposited germanium layer was determined using a quartz thickness gauge. This thickness for different samples was equal to 4, 6, 7, 8, 9, 10, 14, and 18 Å. As a rule, the grown structures were composed of five layers containing germanium quantum dots. The layers with quantum dots were separated by 50-nm-thick undoped silicon layers grown at a temperature of 530°C. The last undoped silicon layer (~100 nm thick) grown at a temperature of 550°C covered the structure with quantum dots. The deposition rates of silicon and germanium atoms were equal to ~0.030 and 0.015 nm/s, respectively. Moreover, we also grew the samples containing one layer with germanium quantum dots that were capped and uncapped with an undoped silicon layer. The pressure in the molecular-beam epitaxy chamber during the removal of the oxide layer and the growth of the structures was raised to $5 \times 10^{-10}$ Torr.

Sizes and concentrations of Ge nanoclusters ($h_{Ge}$ is the effective thickness of the deposited Ge layer, $l$ is the width of the base of Ge nanoclusters, and $h$ is the height of Ge nanoclusters)

| $h_{Ge}$, Å | Size of Ge nanoclusters, nm | | Concentration, $10^{11}$ cm$^{-2}$ |
|---|---|---|---|
| | $l$ | $h$ | |
| 6 | 7–8 | 0.6–1.0 | ~3.5 |
| 8 | 6–15 | 0.6–1.5 | ~6 |
| 10 | 10–15 | 1.0–1.5 | ~5 |
| 14 | 10–15 | 1.0–1.5 | ~2 |

The specially prepared samples were examined using scanning tunneling microscopy. Squares with a side of 8 mm were cut from the initial silicon wafers for examination with the scanning tunneling microscope. After washing and chemical treatment, they were subjected to preliminary heat treatment under the same conditions as the samples for Raman spectroscopic investigations. The atomically clean Si(001) surface was prepared by short-term (2.5 min) annealing at temperatures in the range 900–940°C in the ultrahigh-vacuum molecular-beam epitaxy chamber. Then, a germanium layer was deposited onto the cleaned surface at the same temperature as in the case of the samples for the Raman spectroscopic investigations. The effective thickness of the germanium layer in different samples was equal to 6, 8, 10, and 14 Å. The pressure in the chamber of the scanning tunneling microscope did not exceed $1 \times 10^{-10}$ Torr.

When recording the STM images, the bias voltage $U_t$ was applied to the sample under investigation. The scanning was performed at a constant tunneling current $I_t$.

## 3. RESULTS OF STM INVESTIGATIONS

It was revealed that, at the aforementioned growth temperature, the germanium quantum dots in the form of hut clusters are formed on the Si(001) surface. For the most part, the germanium nanoclusters have a rectangular base. The STM images of the array of hut clusters formed on the Si(001) surface with the effective thickness of the deposited germanium layer $h_{Ge} = 10$ Å are displayed in Fig. 1. The images were obtained at the tunneling current $I_t = 0.1$ nA and the potential difference $U_t = +2.1$ V between the tip of the scanning tunneling microscope and the sample. The STM images of the array of hut clusters formed on the Si(001) surface with the effective thickness of the germanium layer $h_{Ge} = 14$ Å ($U_t = +2.0$ V, $I_t = 0.12$ nA) are displayed in Fig. 2. The concentrations and sizes (heights, base widths) of the germanium quantum dots formed for different effective thicknesses of the germanium layer $h_{Ge}$ are presented in the table. The sizes of germanium nanoclusters increase with an increase in the thickness of the germanium layer and reach a limiting height ($h = 1.0–1.5$ nm) and a limiting width of the base ($l = 10–15$ nm) for an effective thicknesses of the germanium layer $h_{Ge} \sim 10$ Å. At $h_{Ge} > 10$ Å, new germanium hut clusters with considerably smaller sizes are formed between large hut clusters on the free surface of the wetting germanium layer. For $h_{Ge} = 14$ Å, small clusters occupy almost the entire surface of the sample between the large clusters. In this case, the surface density of the large clusters decreases by a factor of almost three as compared to the density of clusters at $h_{Ge} = 10$ Å.

It should be noted that the germanium nanoclusters on the Si(001) surface were examined by scanning tunneling microscopy in our earlier studies.

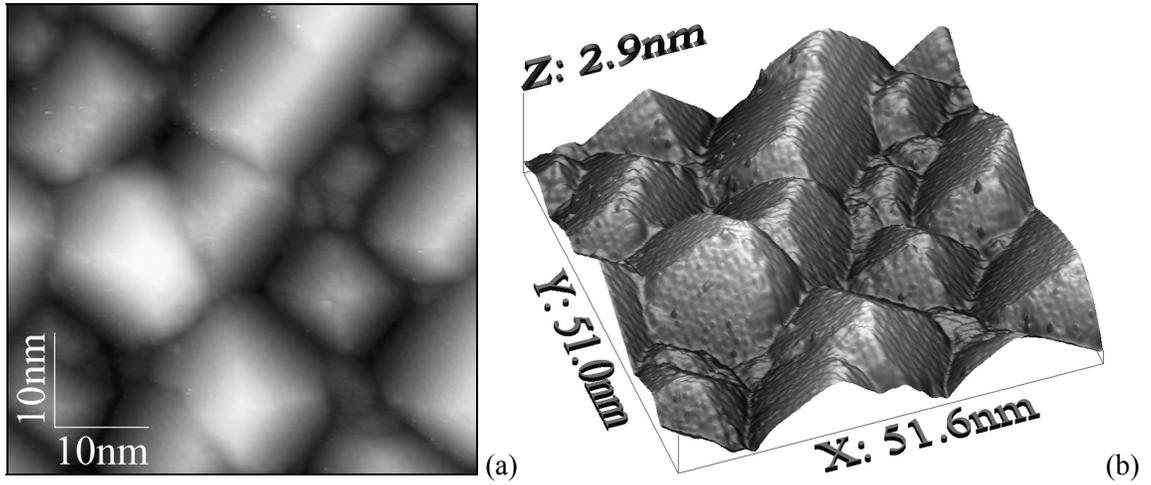

**Fig. 1.** (a) One-dimensional and (b) three-dimensional STM images of the array of Ge hut clusters formed on the Si(001) surface. The effective thickness of the deposited Ge layer is $h_{Ge}$ = 10 Å, the voltage applied to the sample is $U_t$ = +2.1 V, and the tunneling current is $I_t$ = 0.1 nA.

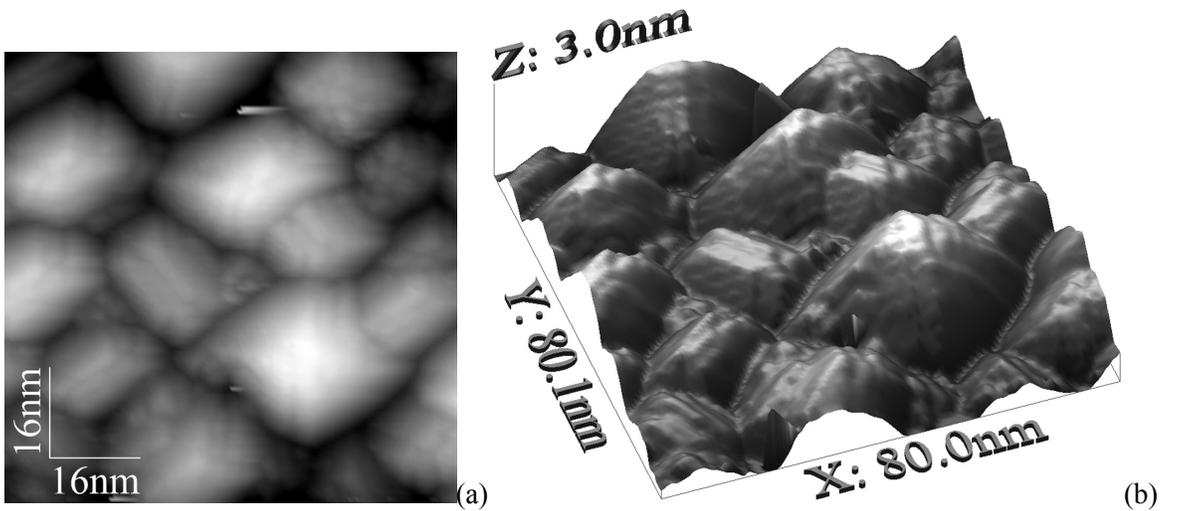

**Fig. 2.** (a) One-dimensional and (b) three-dimensional STM images of the array of Ge hut clusters formed on the Si(001) surface. The effective thickness of the deposited Ge layer is $h_{Ge}$ = 14 Å, the voltage applied to the sample is $U_t$ = +2.0 V, and the tunneling current is $I_t$ = 0.12 nA.

## 4. EXPERIMENTAL RAMAN SPECTRA AND THEIR DISCUSSION

The Raman spectra were recorded on a U-1000 spectrometer in the backscattering geometry upon excitation by an Ar$^{2+}$ laser with the wavelengths λ = 488.0 and 514.5 nm and a He–Cd laser with the wavelength λ = 441.6 nm. The spectral resolution was equal to 1 cm$^{-1}$. The measurements were carried out at the temperature $T$ = 293 K. It is known that Raman spectra of Ge/Si structures with quantum dots contain three dominant peaks: a sharp peak at a frequency of 520 cm$^{-1}$, a peak in the vicinity of 300 cm$^{-1}$, and a band in the vicinity of 400 cm$^{-1}$. These peaks are attributed to vibrations of Si–Si, Ge–Ge, and Ge–Si pairs of neighboring atoms. The frequency of the peak at ω = 520 cm$^{-1}$ is identical to that of optical phonons in bulk silicon; however, the width of the peak in the spectra of Ge/Si nanostructures is considerably larger. The contribution to the Raman spectra in a frequency range of 520 cm$^{-1}$ can be made by scattering from the Si separating layers (50 nm thick), the Si buffer layer, and the GeSi solid solution in the bulk of quantum dots. The contribution to the Raman spectra in the range of Ge–Ge and Ge–Si vibrations can be made by lattice vibrations in the bulk of quantum dots, the wetting layer, and the interface layer at the boundary between the quantum dots and the spacer. The Raman spectra of Ge/Si structures with quantum dots of different heights in the frequency range 225–550 cm$^{-1}$ are depicted in Fig. 3. The dependence of the frequency of the Ge–Ge mode on the thickness of the germanium layer ($h_{Ge}$) is plotted in Fig. 4. It is known that the frequency of the Ramanactive mode at the center of the Brillouin zone in bulk germanium is equal to 301–302 cm$^{-1}$. The shift of the Ge–Ge mode toward the high-frequency range in nanostructures with quantum dots is caused by the elastic compressive stresses in germanium layers in the (001) plane [6–9]. For a germanium thin film laterally compressed so that the parameter of its lattice coincides with the parameter of the silicon substrate lattice, the frequency of the Ge–Ge modes is as high as 319 cm$^{-1}$ [10]. The corresponding frequencies for Ge/Si nanostructures with quantum dots are substantially lower (≈312–314 cm$^{-1}$) [10]. The maximum frequency of the Ge–Ge mode for our samples is equal to 312 cm$^{-1}$. This can be associated with the following factors: the stress in the Ge/Si nanostructures with quantum dots decreases as a result of formation of islands from the film, the quantum dots contain some amount of silicon, and, finally, the size-confinement effects manifest themselves. All three factors can contribute to the spectrum at once. It can be seen from Fig. 4 that the frequency $\omega_{Ge-Ge}$ decreases monotonically with a decrease in the thickness $h_{Ge}$ of the germanium layer in the range 6–10 Å. We believe that, as will be shown below, this decrease is due to the spatial-confinement effect. However, the sample with the thickness $h_{Ge}$ = 14 Å does not obey the above dependence. The frequency of the Ge– Ge mode for this sample is lower than could be expected from the extrapolation of the curve $\omega_{Ge-Ge} = f(h_{Ge})$. According to STM investigations, this sample, apart from large quantum dots, contain quantum dots with a considerably smaller size. It can be seen from the table that the quantum-dot concentration in the sample under consideration is lower than those in other samples by a factor of two. Since the amplitude of resonant Raman scattering is significantly larger than the amplitude of nonresonant Raman scattering, the main contribution to the Raman spectrum is made by the quantum dots that satisfy the resonance conditions. Most likely, the contribution to the resonant scattering for this sample is made by small quantum dots in which the frequency of the Ge–Ge mode is lower. The $E_1$ electronic transitions in small quantum dots correspond better to the resonance conditions. Therefore, their phonon mode manifests itself in the Raman spectra. The line-width of the Ge–Ge and Ge–Si modes is determined by the size distribution of quantum dots and the interaction of the Ge–Ge modes with acoustic phonons 2TA($X$) of silicon, when the frequency of the Ge–Ge mode approaches the frequency $\omega = 300$ cm$^{-1}$. The bandwidths of the Ge–Ge mode are presented in Fig. 4. It can be seen from this figure that the minimum bandwidth is observed for the samples with the thicknesses $h_{Ge} = 9$ and 10 Å. As will be shown below, the Raman scattering in these samples is most similar to resonant scattering. A drastic increase in the bandwidth of the Ge–Ge mode with a frequency of 302 cm$^{-1}$ to $w = 20$ cm$^{-1}$ for the sample with the thickness $h_{Ge} = 6$ Å is most likely explained by the interaction of this mode with the Si 2TA(*X*) phonons at a frequency of ≈300 cm$^{-1}$. The fre quency of the Ge–Si mode does not depend on the germanium layer thickness in the range 6–10 Å and is equal to 419–421 cm$^{-1}$.

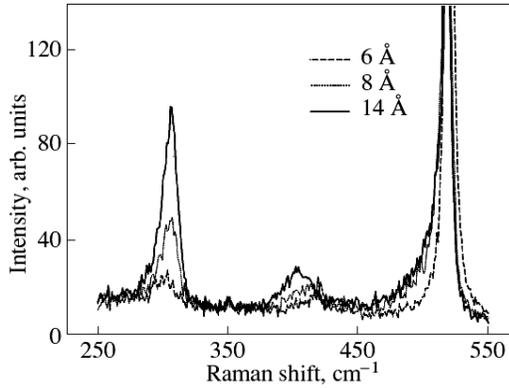

**Fig. 3.** Raman spectra of the Ge/Si samples with Ge quantum dots for the effective thicknesses of the deposited Ge layer $h_{Ge} = 6$, 8, and 14 Å.

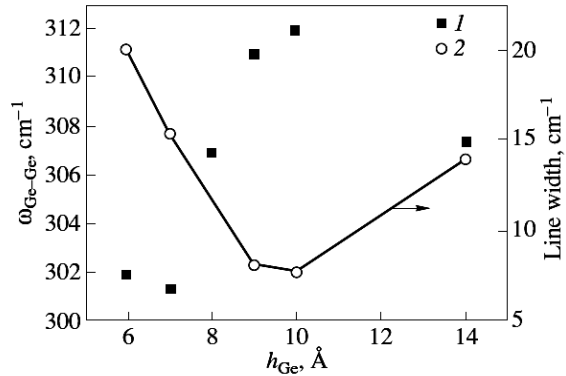

**Fig. 4.** Dependences of (*1*) the frequency and (*2*) the width of the line of the Ge–Ge mode on the effective thickness $h_{Ge}$ of the deposited Ge layer.

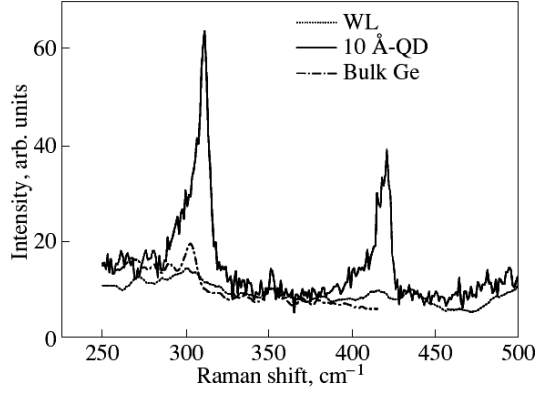

**Fig. 5.** Raman spectra of the Ge/Si structure with quantum dots ($h_{Ge} = 10$ Å), the Ge/Si structure with wetting layers (WLs), and bulk Ge.

In order to determine the influence of the wetting germanium layer on the Raman spectra, we performed the following experiments. We grew the structure that is similar to the structures under investigation and consists of five 4-Å-thick Ge layers separated by silicon layers 50 nm thick. Each Ge layer does not contain any islands and is an analog, to some extent, to the wetting layer in our structures. The thickness of this layer is estimated to be three monolayers (MLs). Figure 5 shows the Raman spectra of Ge/Si structures of two types: one structure contains quantum dots with the thickness $h_{Ge} = 10$ Å, and the other structure involves wetting layers. As can be seen from Fig. 5, the Ge–Si mode weakly manifests itself in the Raman spectrum of the structure with the wetting layers. The intensity of the Ge–Ge line is very low, and its frequency is equal to 302 cm$^{-1}$. Taking into account that, in the structure containing quantum dots at a density of $5 \times 10^{11}$ cm$^{-2}$, the larger part of the surface is occupied by quantum dots and the wetting layer 3 ML thick is located between the quantum dots, the effect of this layer on the Raman spectrum should be weak. This allows the conclusion that the contribution of the wetting layer to vibration spectrum of the Ge/Si structure with quantum dots can be ignored.

It was also of interest to study the influence of the silicon spacer on the Raman spectra. We measured the Raman spectra of the structure containing one 10-Å thick layer of germanium quantum dots uncapped with a silicon layer (Fig. 6). It can be seen from Fig. 6 that the Raman spectrum exhibits only a weak line at a frequency of 302 cm$^{-1}$. Probably, this line is attributed to the Si 2TA(*X*) acoustic phonons. The absence of the line corresponding to the Ge–Si mode indicates that the silicon diffusion from the substrate does not occur at a growth temperature of 350°C. Figure 6 also depicts the Raman spectrum of a similar structure in which the layer of germanium quantum dots is capped by the silicon layer. The deposition of the silicon layer 50 nm thick onto the layer of the germanium quantum dots leads to a radical change in the Raman spectrum. There appear intense lines at frequencies of 308 and 421 cm$^{-1}$. The widths of these lines do not exceed 8 cm$^{-1}$. It follows from this experiment that the layer of germanium quantum dots capped by the silicon layer is strained, as can be judged from the shift of the Ge–Ge mode (308 cm$^{-1}$) by 6 cm$^{-1}$ with respect to the bulk mode (301–302 cm$^{-1}$). The mode at a frequency of 421 cm$^{-1}$ corresponds to the Ge–Si mode in the strained layer of the Ge$_x$Si$_{1-x}$ solid solution. The frequency of the Ge–Si mode in the Ge$_{0.65}$Si$_{0.35}$ bulk alloy is equal to 406 cm$^{-1}$ [3]. The shift of this mode in the spectrum of the structure with quantum dots by 15 cm$^{-1}$ is associated with the stresses in the layers with quantum dots. Therefore, silicon that caps the layer of germanium quantum dots induces stresses in the layer with quantum dots and diffuses into the bulk of quantum dots with the formation of the Ge$_x$Si$_{1-x}$ solid solution. The intense diffusion through the surface between the layer with quantum dots and the capped silicon layer is explained by the considerable inhomogeneity of the surface in both the growth and longitudinal directions. Microinhomogeneities (quantum dots) produce large composition and elastic stress gradients that are responsible for the diffusion. Moreover, the area of this surface in view of roughness is larger than that of the surface adjacent to the substrate. The silicon concentration in the volume of quantum dots in this sample was estimated to be 38% from the intensity ratio of the peaks corresponding to the Ge–Ge and Ge–Si modes. The calculation technique will be described below. An increase in the intensity of the lines of the Ge–Ge and Ge–Si modes in the capped structure as compared to the uncapped structure is explained by the increase in the energy of the gaps $E_1$ and $E_1 + \Delta_1$ for the Ge$_x$Si$_{1-x}$ alloy. In Ge$_x$Si$_{1-x}$ bulk alloys, the energy gap at the $E_1$ point varies from 2.2 eV ($x = 1$) to 2.77 eV ($x = 0.50$) [11, 12]. In our sample with $x = 0.62$, the transition energy is $E_1 \approx 2.6$ eV. We believe that the excitation of the structures under investigation by the Ar$^+$ laser with a wavelength of 488 nm ($E = 2.54$ eV) provides the resonance condition for the interaction of the $E_1$ excitons in the quantum dots with the Ge–Ge and Ge–Si modes. This results in a considerable increase in the scattering amplitude and a decrease in the bandwidth of the above modes. For comparison, the Raman spectrum of the bulk germanium sample is depicted in Fig. 5. It is evident that the amplitude of resonant Raman scattering in the Ge/Si structure with quantum dots is substantially higher than the corresponding amplitude for bulk germanium. Similar results were obtained by Talochkin et al. [13].

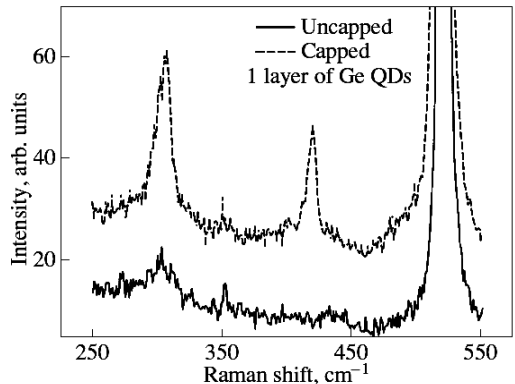

**Fig. 6.** Raman spectra of the Ge/Si structures with one layer of Ge quantum dots ($h_{Ge} = 10$ Å) uncapped and capped with the Si layer.

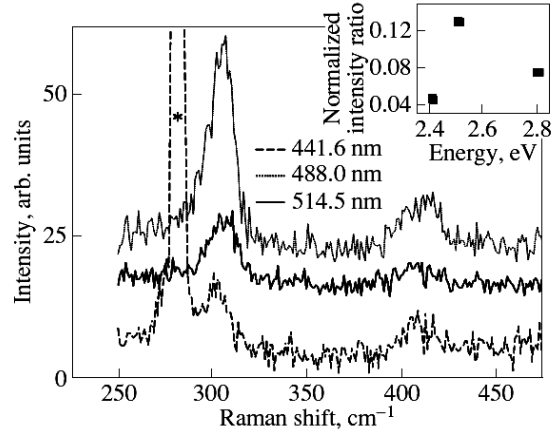

**Fig. 7.** Raman spectra of the Ge/Si structure with the effective thickness of the deposited Ge layer $h_{Ge}$ = 8 Å upon excitation by the Ar laser with the wavelengths λ = 488.0 and 514.5 nm and the He–Cd laser with the wavelength λ = 441.6 nm. The asterisk indicates the discharge line of the He–Cd laser (282 cm$^{-1}$). The inset shows the dependence of the intensity of the line of the Ge–Ge mode (normalized to the intensity of the line of the Si–Si mode) on the energy of the exiting laser.

We also investigated the Raman spectra at different excitation energies, namely, upon excitation by the Ar$^+$ laser with the wavelengths λ = 488.0 and 514.5 nm and the He–Cd laser with the wavelength λ = 441.6 nm. These spectra for the structure with the thickness $h_{Ge}$ = 8 Å are shown in Fig. 7. It can be seen from this figure that the intensity of the line corresponding to the Ge–Ge mode is maximum upon excitation by the laser with a wavelength of 488 nm. The intensity of the Ge–Ge mode was normalized to the intensity of the Si–Si mode. The results obtained are presented in the inset to Fig. 7. The resonant character of Raman scattering in Ge/Si structures with germanium quantum dots was previously investigated by Kwook et al. [7] upon excitation by a tunable laser in the excitation energy range 2.0–2.7 eV. It was demonstrated that the maximum intensity of the Ge–Ge mode corresponds to an energy of 2.43 eV. Our experimental results are in agreement with the data obtained in the above work. The difference is that, in our case, the maximum intensity of the line of the Ge–Ge mode corresponds to the energy $E$ = 2.5 eV. Possibly, this is associated with the fact that the sizes of quantum dots in our structures are smaller by a factor of approximately 1.5.

The frequency shift of this mode is noteworthy. In particular, we have $\omega_{Ge-Ge}$ = 308.6 cm$^{-1}$ at λ = 514.5 nm, $\omega_{Ge\,Ge}$ = 307.5 cm$^{-1}$ at λ = 488.0 nm, and $\omega_{Ge-Ge}$ = 302.9 cm$^{-1}$ at λ = 441.6 nm. It can be seen that an increase in the energy of the exciting laser leads to a decrease in the frequency of the Ge–Ge mode. Our data correlate with the results obtained by Milekhin et al. [14], who demonstrated that, in the excitation energy range 2.0–2.7 eV, the frequency $\omega_{Ge-Ge}$ noticeably decreases beginning with energies $E$ > 2.5 eV. We explain this finding by using the dependence of the frequency $\omega_{Ge-Ge}$ on the germanium layer thickness (Fig. 4) and the results of STM investigations (Figs. 1, 2) according to which the quantum dots in the germanium layers are characterized by a size dispersion. At $E$ = 2.8 eV (λ = 441.6 nm), small quantum dots having a lower phonon frequency make a resonant contribution to the Raman spectrum. The line of the Si–Si mode in the bulk of quantum dots is superimposed on a very intense Raman peak associated with the spacer layers, the buffer layer, and the substrate. However, the manifestation of this line can be revealed from an increase in the scattering intensity of the low-energy wing of the line (Fig. 6) as compared, for example, with the scattering from the silicon substrate.

The germanium concentration in quantum dots can be estimated by comparing the integrated intensities of the Ge–Ge and Ge–Si lines and using the formula $I_{Ge-Ge}/I_{Ge-Si} = Bx/2(1 - x)$ [4], where $x$ is the Ge concentration. The above coefficient is $B = 1$ for bulk GeSi alloys [4] and $B = 3.2$ for unstrained epitaxial layers $Ge_xSi_{1-x}$ [8, 10]. According to Volodin et al. [6], this coefficient is $B \approx 2$ for the Ge/Si structure with germanium quantum dots.

By using the data obtained in [6], we calculated the germanium concentration also under the assumption that $B = 2$. The dependence of the germanium concentration $x$ in the quantum dots on the thickness of the deposited germanium layer according to calculations at $B = 2$ is plotted in Fig. 8. The germanium concentration was most accurately determined in the samples with the thicknesses $h_{Ge} = 9$ and 10 Å, for which the bandwidth of the Ge–Ge and Ge–Si modes is minimum. The germanium concentrations obtained are averaged over the volume of quantum dots, because germanium is non-uniformly distributed over the volume of quantum dots [15]. As can be seen from Fig. 8, the silicon concentration remains unchanged within the limits of experimental error in the range of germanium layer thicknesses 6–10 Å and is equal to $34 \pm 2\%$. However, it follows from Fig. 4 that the frequency of the Ge–Ge mode decreases monotonically from 312 to 301 cm$^{-1}$ in the aforementioned thickness range. We considered three factors that can be responsible for the decrease in the frequency of the Ge–Ge mode with a decrease in the germanium layer thickness: the increase in the silicon concentration in the quantum dots, the decrease in the stresses, and the spatial-confinement effect. The calculations of the stresses show that the strain does not depend on the sizes of quantum dots when their shape is retained [16]. Consequently, the first two factors should be excluded. Therefore, the decrease in the frequency of the Ge–Ge mode should be attributed to the effect of spatial confinement on the frequency of the phonon mode due to the negative dispersion of the longitudinal optical (LO) mode of germanium. This effect noticeably manifests itself for effective thicknesses $h_{Ge} \leq 10$ Å. In order to confirm our assumptions, the confinement effect was evaluated by approximating the dispersion curve of Ge LO phonons in the range of wave vectors $q/q_{max} = 0$–1/2 along the [100] direction [17] with the use of the relationship

$$\omega = \omega_0[1 - \alpha(q/q_{max})^2],$$

where $q_{max} = \pi/a_{Ge}$ and $a_{Ge}$ is the lattice parameter of germanium. As a result, we obtained $\alpha = 8/15$. By using our data $\omega_0 = 312$ cm$^{-1}$ and the frequency shift $\Delta\omega = -10$ cm$^{-1}$, we obtain $q/q_{max} = 0.245$. Setting $q = \pi/d$ (where $d$ is the size of the confinement region), we find $d = 23$ Å. This value is close to the side of $\approx 24.3$ Å of the cube with the volume equal to the volume of the pyramid with $h = 6$ Å, $l_1 = 60$ Å, and $l_2 = 120$ Å.

The confinement effect also manifests itself in the asymmetric shape of the lines of the Ge–Ge and Ge–Si modes (Figs. 5, 6), which contain an extended tail in the low-frequency range and a sharp edge in the high-frequency range.

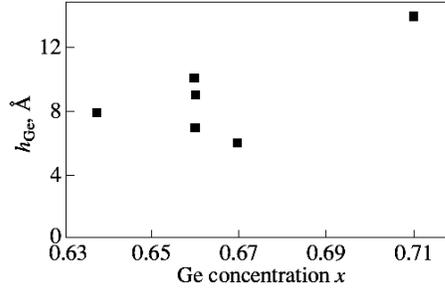

**Fig. 8.** Dependence of the Ge concentration $x$ in quantum dots on the effective thickness $h_{Ge}$ of the deposited Ge layer.

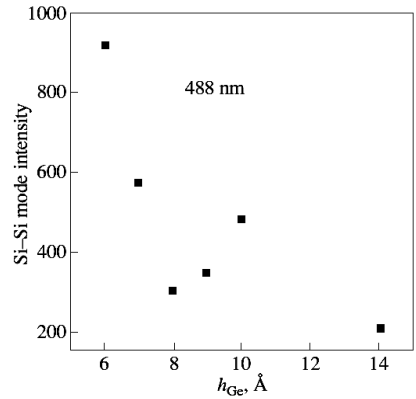

**Fig. 9.** Dependence of the intensity of the line of the Si–Si mode on the effective thickness of the deposited Ge layer upon excitation by the Ar$^+$-laser with the wavelength $\lambda = 488$ nm.

We analyzed the dependence of the intensity of the line of the Si–Si mode on the germanium layer thickness upon excitation by the Ar$^+$ laser with the wavelength $\lambda = 488$ nm (Fig. 9). As can be seen from Fig. 9, the line intensity at $h_{Ge} = 10$ Å is two times higher than that at $h_{Ge} = 14$ Å. This is most likely associated with the decrease in the absorption of exciting light in the germanium layers with the decrease in the germanium layer thickness. According to the estimates made in [13], the absorption coefficient at the energy $E = 2.54$ eV is approximately equal to $10^6$ cm$^{-1}$ for germanium quantum dots and $5 \times 10^5$ cm$^{-1}$ for bulk germanium. However, a further decrease in the germanium layer thickness (in the range 10–8 Å) leads to a considerable decrease in the intensity of the line of the Si–Si mode. This non-monotonic behavior of the intensity of the line of the Si–Si mode with the decrease in the quantity $h_{Ge}$ can be explained by the influence of the size confinement on the exciton transition energy $E_1$, which increases with a decrease in the size of germanium quantum dots. As the effective thickness $h_{Ge}$ decreases, the corresponding absorption should decrease monotonically. However, with the decrease in the thickness $h_{Ge}$ to 8 Å, the energy $E_1$ most likely passes through a resonance with the excitation energy. As a consequence, the absorption in the germanium layer increases and, hence, the intensity of scattering from the Si–Si mode decreases. The further decrease in the germanium layer thickness ($h_{Ge}$ = 7, 6 Å) is accompanied by a drastic increase in the intensity of scattering from the Si–Si mode, which indicates a decrease in the absorption in the germanium layers.

## 5. CONCLUSIONS

Thus, in this study, we analyzed the role of different factors, such as the interdiffusion, elastic stresses, and spatial confinement, in the formation of the Raman spectra of Ge/Si structures with germanium quantum dots. It was demonstrated that silicon diffuses into germanium quantum dots from the capped silicon layer rather from the underlying layer through faces and edges of quantum dots, where elastic strains and composition gradients are maximum.

It was revealed that a decrease in the sizes of quantum dots in the range 10–6 Å results in a decrease in the frequency of the Ge–Ge mode by 10 cm$^{-1}$ due to the effect of spatial confinement on phonons. The Raman spectra of the samples with the effective thicknesses $h_{Ge}$ = 10 and 9 Å at a silicon concentration of 35% in quantum dots have a resonance character with respect to the Ge–Ge and Ge–Si modes. The dependence of the intensity of the line of the Si–Si mode on the germanium layer thickness upon excitation by the Ar$^+$ laser with an energy of 2.54 eV exhibits a non-monotonic behavior: the intensity of the line is minimum for the samples characterized by resonant scattering of the Ge– Ge and Ge–Si vibrational modes and maximum for the samples with the minimum Ge layer thickness (7, 6 Å). This is explained by the effect of size confinement on the electronic spectrum of $E_1$ excitons.

The STM images were prepared for publication with the WSxM free software for scanning probe microscopy [18].

## ACKNOWLEDGMENTS

This study was supported by the Russian Foundation for Basic Research (project no. 07-02-00899-a), the Presidium of the Russian Academy of Sciences (Program "Quantum Nanostructures," project no. 5.4), and Federal Agency for Science and Innovation of the Ministry of Education and Science of the Russian Federation (State Contract no. 02.513.11.3130).